\documentclass[pdflatex,sn-nature]{sn-jnl}

\usepackage[T1]{fontenc}
\usepackage{graphicx}
\usepackage{amsmath,amssymb,amsfonts}
\usepackage{bm}
\usepackage{siunitx}

\graphicspath{{figures/}}

\begin{document}

\title[Translation-to-rotation conversion of skyrmion clusters]
{Skyrmion gears: translation-to-rotation conversion of skyrmion clusters along periodic substrates}

\author*[1]{\fnm{N. P.} \sur{Vizarim}}
\email{nicolas.vizarim@unesp.br}

\author[2]{\fnm{C. J. O.} \sur{Reichhardt}}

\author[2]{\fnm{C.} \sur{Reichhardt}}

\affil*[1]{
\orgdiv{Department of Physics and Meteorology, School of Sciences},
\orgname{S\~ao Paulo State University},
\orgaddress{
\city{Bauru},
\state{S\~ao Paulo},
\country{Brazil}
}
}

\affil[2]{
\orgdiv{Theoretical Division and Center for Nonlinear Studies},
\orgname{Los Alamos National Laboratory},
\orgaddress{
\city{Los Alamos},
\state{New Mexico},
\postcode{87545},
\country{USA}
}
}

\abstract{
We investigate how the translational motion of a compact skyrmion
cluster can be converted into collective rotation through interactions
with a periodic obstacle substrate. Using an effective particle model,
we consider a seven-skyrmion cluster driven along a one-dimensional
array of repulsive obstacle sites. For suitable obstacle spacings and
interaction strengths, repeated interactions with the substrate produce
coherent clockwise or counterclockwise rotation, giving rise to
gear-like cluster dynamics. We characterize these states using the
translational velocity, Hall angle, angular velocity, structural
deformation, and a dimensionless gear ratio that relates cluster
rotation to motion over the substrate periodicity. Dynamical phase
diagrams identify rotating, nonrotating, wall-guided, transverse, and
deformed regimes. We further show that coherent rotation persists in
the overdamped limit, demonstrating that the conversion mechanism does
not require an intrinsic Magnus force and may therefore extend to
other driven cluster systems, including colloidal assemblies. These
results provide a simple route for controlling collective rotation
through structured substrates.
}

\keywords{
magnetic skyrmions,
collective dynamics,
periodic substrate,
rotational motion,
driven particle systems,
skyrmion Hall effect
}

\maketitle

\section{Introduction}
\label{sec:introduction}

A wide range of particle-like systems can form cluster states,
including soft matter such as colloids
\cite{Yang12,Ducrot17,Fang25} and particles on fluid interfaces
\cite{Grzybowski00,Hooshanginejad24}.
Cluster formation can also occur in particle-like hard condensed matter
systems, such as magnetic skyrmions
\cite{Leonov2016,Kim2020,Rozsa16,Du18,Loudon18,Yu20,Tang2021,Wang22,Leonov23,zhang_stable_2024,Vizarim26},
vortices in multiband superconductors
\cite{Chaves11,Gutierrez12,Varney13,Brems22},
or mixed superconducting-magnetic systems \cite{Neto22}.
In certain cases, such as for magnetic skyrmions, a cluster can be set
into motion under an external driving force, making possible the
experimental observation of both the motion and rotation of a cluster
\cite{Yu20,Wang22}.
In other systems, such as colloidal assemblies, clusters can be moved
by external fluid flow or an electric field \cite{Cao21}.

When lattice-forming particles, such as vortices in superconductors,
magnetic skyrmions, or colloids, are driven over a random or periodic
landscape, they can exhibit transitions among a number of distinct
dynamical states, such as a pinned state, disordered flow, soliton-like
flows, and different types of smectic and crystalline flow states
\cite{Reichhardt98,MacDonald03,Bohlein12,Tierno12a,Reichhardt17,Stoop18}.
Driven magnetic skyrmions can experience an additional Magnus force
that affects how the skyrmions interact with pinning and induces a
skyrmion Hall effect
\cite{Nagaosa13,Lin13,Jiang17,Litzius17,Vizarim20,Reichhardt22a}.
Artificial arrays and interfaces can be used to pin or repel skyrmions, and
there has been considerable work on understanding how to control the
motion of skyrmions using structured defect arrays
\cite{Reichhardt15a,Castell-Queralt19,Fernandes20,Feilhauer20,
Vizarim21,Juge21,Gruber22,Reichhardt22a,Vizarim22}.

Despite the number of systems in which cluster states can appear,
only a limited number of studies have been performed on the depinning
and driven dynamics of clusters on disordered and ordered substrates.
These have demonstrated directional locking and a transition in which
clusters can break up and reform at high drives
\cite{Reichhardt03,Drocco13,Cao21}.
A distinctive feature that separates cluster motion from that of
single particles is that the cluster can have internal degrees of
freedom, such as the ability to distort, as well as rotational degrees
of freedom.
An open question is how cluster rotation couples to translational
motion on periodic substrates, particularly when the interparticle
spacing within the cluster is commensurate with the substrate
periodicity.
If a cluster is moving over an ordered lattice, it should be possible
to find a dynamical regime in which the cluster also exhibits coherent
rotational dynamics, such as when the spacing of the particles in the
cluster matches the substrate lattice spacing.
If the spacing does not match or is incommensurate, the cluster could
effectively slide over the substrate instead of rotating, or a
transition could occur from rolling to sliding motion.

Here we consider the driven dynamics of a seven-particle skyrmion
cluster subjected to both damping and Magnus terms and interacting
with a line of obstacle sites.
When we apply a dc drive parallel to the obstacle line, the skyrmion
Hall effect deflects the cluster toward the obstacles.
As a function of increasing drive, we identify several distinct
dynamical regimes.
At low drives, the cluster can become commensurate with the obstacle
array and roll along the interface as individual skyrmions are
temporarily trapped.
We characterize this motion using the dimensionless gear ratio,
$G_6=6a\langle\omega\rangle/(2\pi\langle V_x\rangle)$.
The rotation is the most pronounced when the interparticle spacing
within the cluster matches the periodicity of the obstacle array, and
counterclockwise rotational motion can also occur under certain
conditions.
At intermediate drives, the cluster can transition to a sliding state
in which it aligns with the center of the obstacle line and moves along
it without rotation.
At higher drives, the cluster crosses the obstacle line and then
travels at a finite Hall angle, with strong distortions induced by the
obstacles at the highest drives.
We map out these regimes as a function of the drive, obstacle lattice
spacing, and ratio of the Magnus coefficient to the damping
coefficient.
Finally, we show that rotation persists even in the absence of a
Magnus force, suggesting that similar behavior may arise in other
cluster-forming systems with repulsive obstacle interactions.

\begin{figure}[t]
  \centering
  \includegraphics[width=0.70\linewidth]{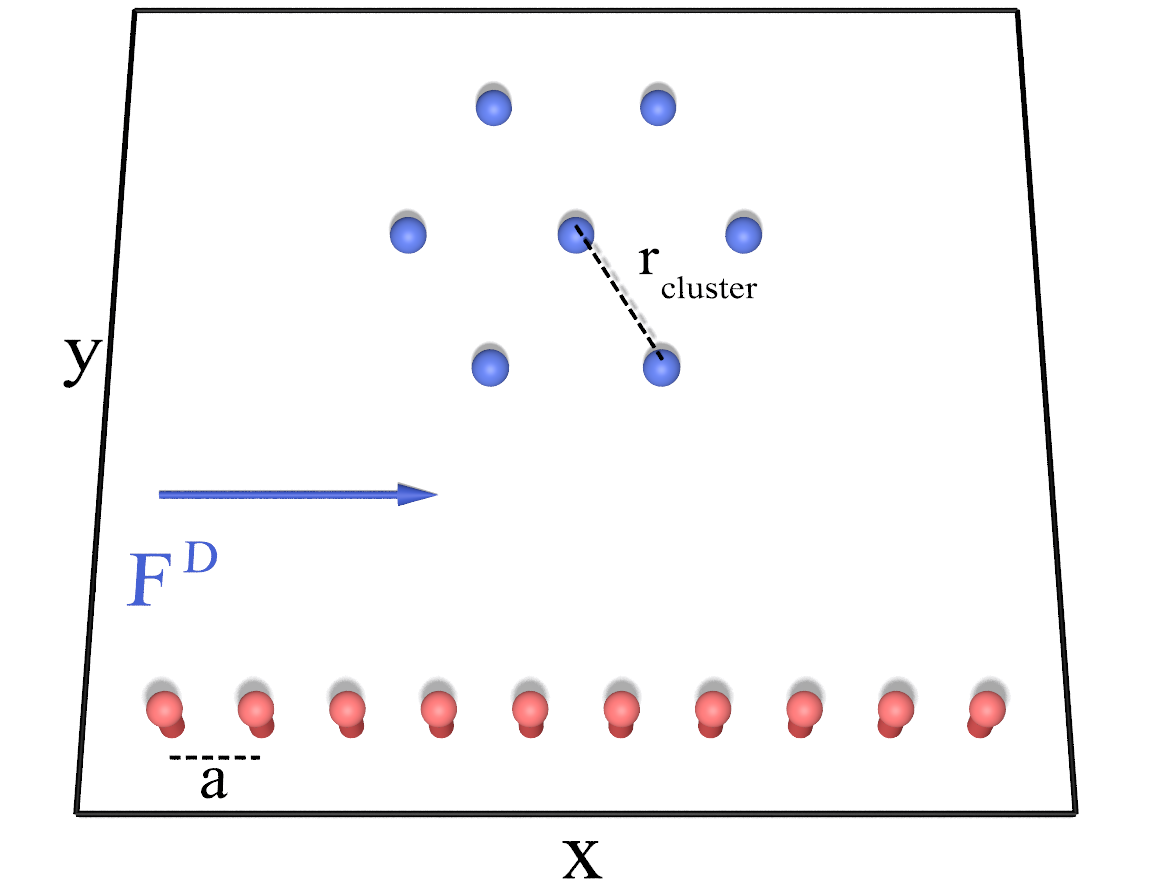}
  \caption{
  Schematic of the simulated system.
  A seven-skyrmion cluster (blue spheres), composed of one central
  skyrmion caged by six surrounding skyrmions, interacts with a
  one-dimensional array of repulsive obstacle sites (red posts)
  located along $y=2.0$.
  The obstacle sites are equally spaced by a distance $a$.
  The quantity
  $r_{\rm cluster}\approx 2^{1/6}\sigma$ denotes the characteristic
  distance between the central skyrmion and the surrounding particles
  in the relaxed cluster configuration.
  An external dc drive $\mathbf{F}^{D}$ is applied along the positive
  $x$ direction unless otherwise noted.
  Periodic boundary conditions are applied in both the $x$ and $y$
  directions.}
  \label{fig1}
\end{figure}

Figure~\ref{fig1} illustrates the geometry of the model used throughout
this work.

\section{Results}
\label{sec:results}

\subsection{Cluster dynamics}
\label{subsec:principal-result}

\begin{figure}[t]
  \centering
  \includegraphics[width=0.8\linewidth]{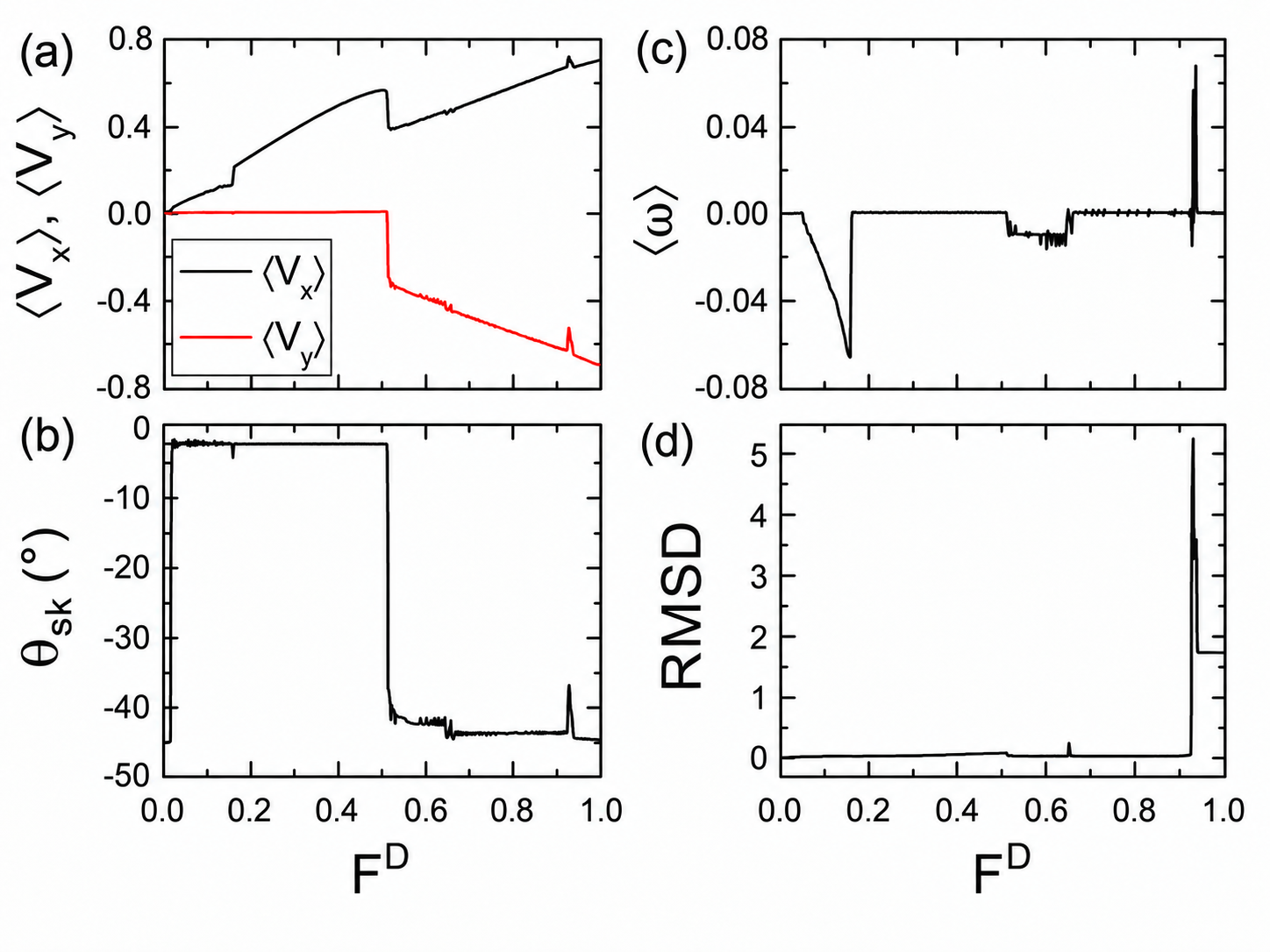}
  \caption{
  Driven dynamics of the seven-skyrmion cluster for obstacle spacing
  $a=1.0$ and interaction strength $\epsilon=0.3$.
  (a) Average longitudinal and transverse velocity components,
  $\langle V_x\rangle$ (black) and $\langle V_y\rangle$ (red), vs
  applied drive $F^D$.
  (b) The corresponding skyrmion Hall angle $\theta_{\rm sk}$ vs
  $F^D$.
  (c) The average angular velocity $\langle\omega\rangle$ of the
  cluster vs $F^D$.
  (d) The root-mean-square deviation (RMSD) from the relaxed cluster
  configuration vs $F^D$.}
  \label{fig2}
\end{figure}

We first consider a representative case with obstacle spacing
$a=1.0$ and skyrmion interaction strength $\epsilon=0.3$.
For this choice of parameters, the cluster exhibits several distinct
dynamical regimes as the drive is increased.
In Fig.~\ref{fig2}(a) we plot the average longitudinal and transverse
velocity components, $\langle V_x\rangle$ and $\langle V_y\rangle$,
versus $F^D$.
At low drives, the motion is predominantly along the $x$ direction,
with $\langle V_y\rangle\approx0$.
In this regime, the periodic obstacle array guides the cluster parallel
to the obstacle line and strongly suppresses the transverse motion
produced by the Magnus term.
For $F^{D}<0.162$, $\langle V_x\rangle$ increases with drive and
displays bumps and variations associated with changes in the cluster
motion relative to the obstacle sites.
For $0.162<F^{D}\lesssim0.500$ the dynamics is relatively stable and
$\langle V_x\rangle$ increases smoothly, while the transverse velocity
remains very small, $\langle V_y\rangle\approx0$.
For $F^D\gtrsim0.5$, a sharp change occurs:
$\langle V_y\rangle$ becomes finite and negative, while
$\langle V_x\rangle$ decreases abruptly.
Here the external drive supplies enough energy for the cluster to
cross the obstacle wall and move in a direction very close to that of
the intrinsic skyrmion Hall angle,
$\theta_{\rm sk}^{\rm int}=-45^{\circ}$, as shown in
Fig.~\ref{fig2}(b).
For $F^{D}<0.5$, the cluster instead moves along the obstacle wall
with $\theta_{\rm sk}=0^{\circ}$.

The rotational dynamics is shown in Fig.~\ref{fig2}(c), where we plot
$\langle \omega\rangle$ versus $F^D$.
At low drives, for $F^D<0.162$, the cluster develops a finite negative
angular velocity corresponding to clockwise rotation.
The magnitude of $\langle\omega\rangle$ increases with increasing
drive before the rotational motion terminates for $F^D>0.162$, where
$\langle\omega\rangle$ drops nearly to zero.
For intermediate drives, the cluster moves with little or no net
rotation.
A weaker rotational response reappears approximately in the interval
$0.5\lesssim F^D\lesssim0.65$, coincident with the transition from
guided to finite-Hall-angle motion shown in Fig.~\ref{fig2}(b),
followed by a return of $\langle\omega\rangle$ to nearly zero at
higher drives.

Figure~\ref{fig2}(d) shows the RMSD relative to the relaxed cluster
configuration.
The RMSD remains small throughout the low and intermediate drive
regimes, demonstrating that the cluster retains approximately the
same internal structure while undergoing both translational and
rotational motion.
A pronounced increase in the RMSD occurs near $F^D\approx0.93$, a
drive at which there are also sharp features in the velocity, Hall
angle, and angular velocity signatures.
At this drive, the seven-skyrmion cluster becomes strongly distorted
and the sixfold symmetry is eventually lost.
The corresponding large values of $\langle\omega\rangle$ therefore do
not represent a coherent rotation of an intact cluster.

\begin{figure}[t]
  \centering
  \includegraphics[width=1.00\linewidth]{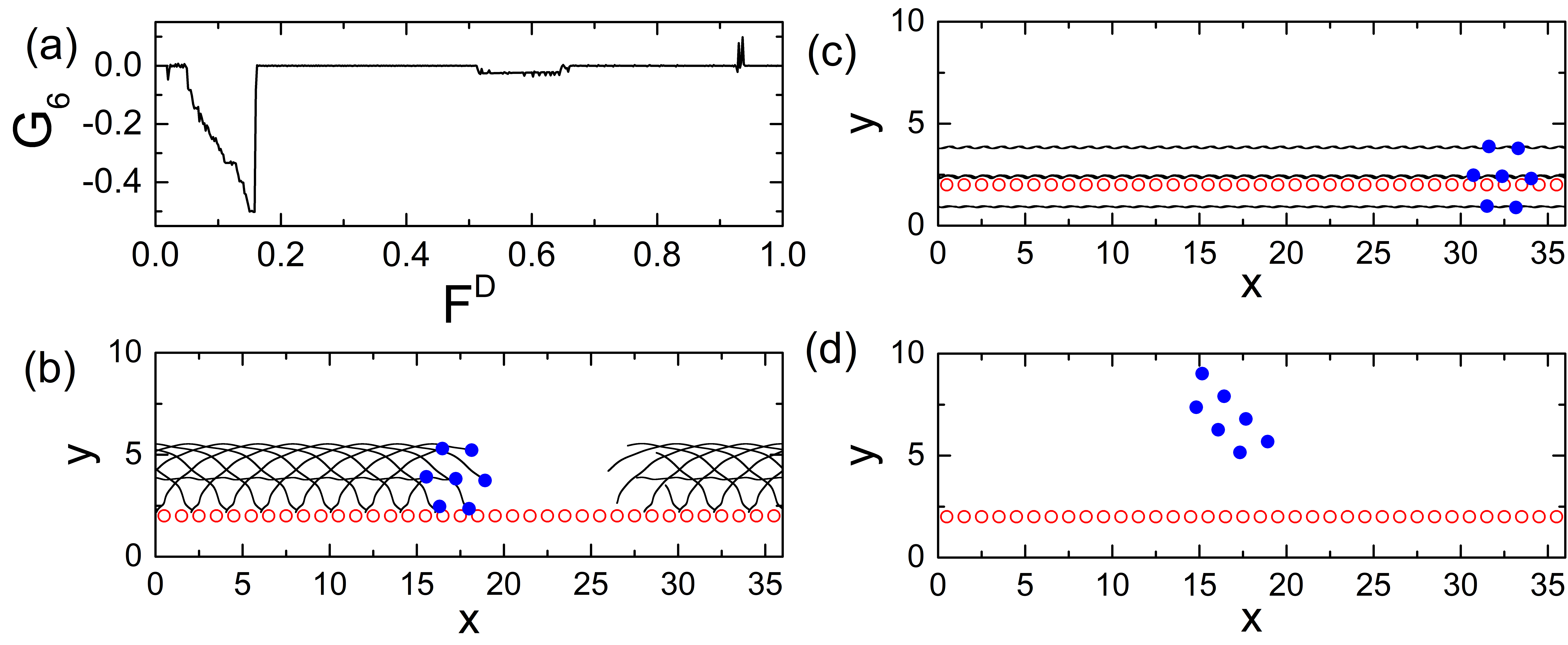}
  \caption{
  Rotational response and representative particle trajectories for
  $a=1.0$ and $\epsilon=0.3$.
  (a) Gear ratio $G_6$ vs applied drive $F^{D}$.
  (b--d) Images showing the repulsive obstacle sites (red circles), the
  positions of the particles (blue circles), and the particle
  trajectories (black lines).
  (b) $F^{D}=0.156$, corresponding to the regime of gear-like rotation
  with $G_6\approx-1/2$.
  (c) $F^{D}=0.3$, where the rotational response is suppressed.
  (d) Snapshot at $F^{D}=1.0$, illustrating the distorted cluster
  configuration that appears above the high-drive increase in the
  RMSD.}
  \label{fig3}
\end{figure}

Figure~\ref{fig3} provides a more direct view of the relation between
the translational and rotational motion of the cluster for the same
system shown in Fig.~\ref{fig2}.
In Fig.~\ref{fig3}(a) we plot the gear ratio $G_6$ versus applied
drive $F^D$.
At low drives, the cluster develops a pronounced rotational response,
with $G_6$ reaching approximately $-1/2$ before the rotation is
abruptly suppressed near $F^D\simeq0.16$.
Since $G_6$ measures the cluster rotation in units of $60^\circ$ per
obstacle lattice period, $|G_6|=1/2$ corresponds to one $60^\circ$
rotation for every two obstacle periods traversed by the cluster, or
equivalently an average rotation of $30^\circ$ per obstacle period.
The negative sign indicates clockwise rotation for motion along the
positive $x$ direction.

The particle trajectories are illustrated in Fig.~\ref{fig3}(b) at
$F^D=0.156$, where $G_6\approx-1/2$.
Here the individual trajectories form periodic curved structures as
the cluster travels along the periodic obstacle array.
The repeated interaction with the repulsive obstacle sites produces a
sequence of transverse deflections that collectively generates the
rotation of the cluster.
In contrast, in Fig.~\ref{fig3}(c) at $F^D=0.3$, where
$G_6\simeq0$, the trajectories become nearly periodic and
predominantly translational.
In this regime, the skyrmions undergo small oscillatory deflections as
they move along the obstacle line, but these motions are too small to
induce a net rotation of the cluster.
At larger drives, the sharp spike in $G_6$ near $F^D\approx0.93$
coincides with a strong increase in the cluster deformation, as also
indicated by the RMSD in Fig.~\ref{fig2}.
A representative snapshot at $F^D=1.0$ in Fig.~\ref{fig3}(d) shows
that the cluster has lost its sixfold symmetry and has adopted a
distorted configuration.

\subsection{Obstacle-site spacing and interaction strength}
\label{subsec:robustness}

\begin{figure}[t]
  \centering
  \includegraphics[width=1.0\linewidth]{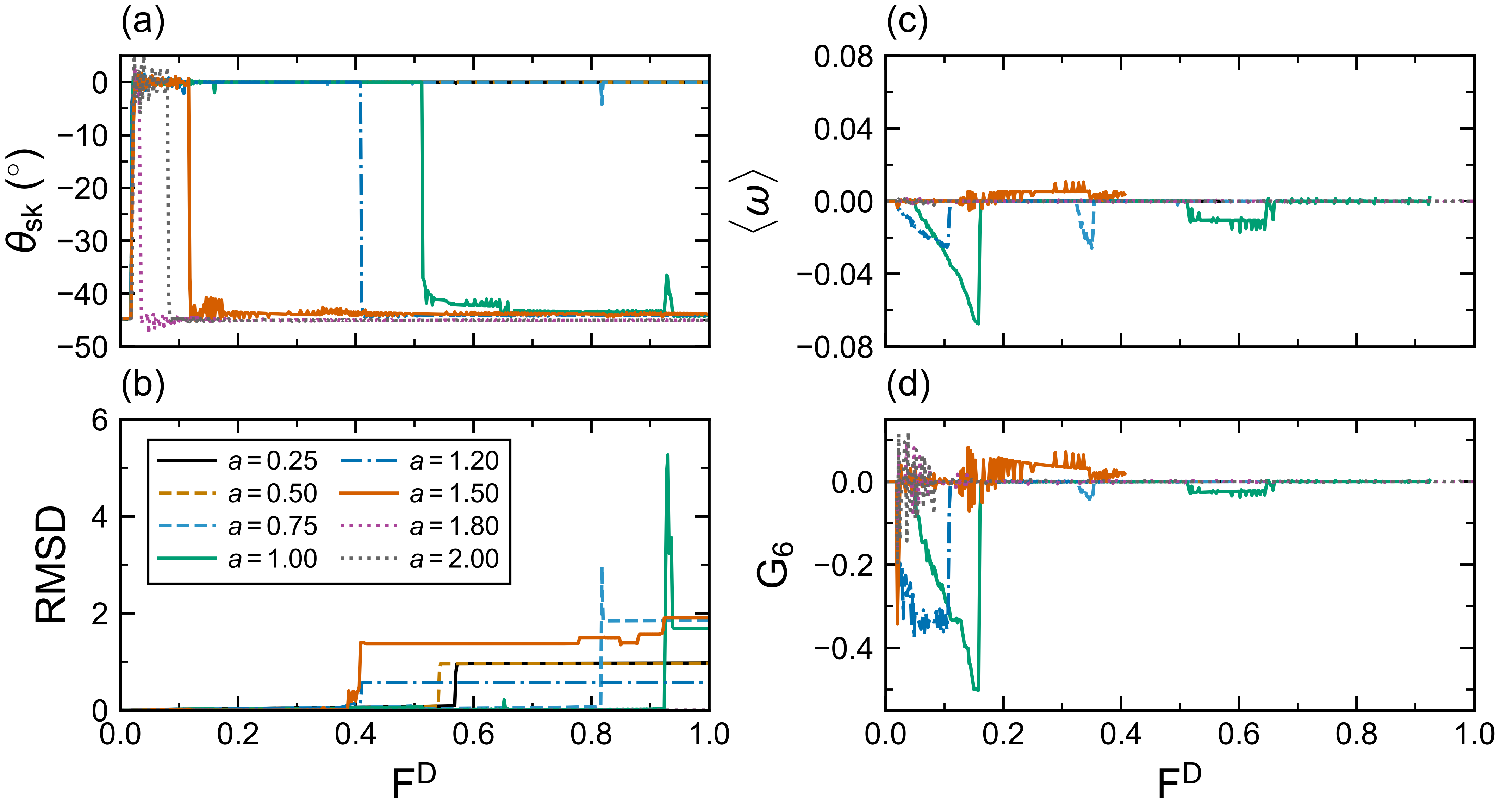}
  \caption{
  Dependence of the cluster dynamics on the obstacle-site spacing $a$
for fixed interaction strength $\epsilon=0.3$.
(a) Skyrmion Hall angle $\theta_{\rm sk}$ versus $F^D$.
(b) Root-mean-square deviation (RMSD) from the relaxed cluster
configuration versus $F^D$.
(c) Average angular velocity $\langle \omega \rangle$ versus $F^D$.
(d) Gear ratio $G_6$ versus $F^D$.
The different curves correspond to
$a=0.25$ (black solid),
$0.50$ (gold dashed),
$0.75$ (light-blue dashed),
$1.00$ (green solid),
$1.20$ (blue dash-dotted),
$1.50$ (orange solid),
$1.80$ (magenta dotted), and
$2.00$ (gray dotted).}
  \label{fig4}
\end{figure}

Now that we have established the microscopic origin of the gear-like
dynamics for a representative system, we examine the effect of the
obstacle-site spacing $a$ on the rotational response at fixed
interaction strength $\epsilon=0.3$.
Figure~\ref{fig4} shows the skyrmion Hall angle
$\theta_{\mathrm{sk}}$, the root-mean-square deviation (RMSD), the
mean angular velocity $\langle\omega\rangle$, and the dimensionless
gear ratio $G_6$ versus $F^D$.

The Hall response in Fig.~\ref{fig4}(a) depends strongly on the
obstacle-site spacing.
Several of the curves exhibit an extended interval with
$\theta_{\mathrm{sk}}\simeq0$, followed by a relatively abrupt
transition to a negative Hall angle of
$\theta_{\mathrm{sk}}\approx-45^\circ$.
The transition shifts to lower values of $F^D$ as $a$ increases.
This behavior is consistent with a change in the effective
permeability of the obstacle barrier.
Closely spaced obstacle sites can form a more continuous barrier that
inhibits transverse motion, favoring transport along $+x$ over an
extended drive interval.
Increasing $a$ opens wider gaps between neighboring obstacle sites,
facilitating passage across the barrier and allowing a finite negative
Hall response to develop at lower drives.

The plot of RMSD versus $F^D$ in Fig.~\ref{fig4}(b) shows that the
structural response also depends on $a$.
For several obstacle spacings, the RMSD remains close to zero over a finite
drive interval before increasing to a nonzero plateau.
Additional steps and isolated peaks occur at larger drives.
These features indicate modifications from the reference cluster
configuration and show that the obstacle spacing has a strong influence
on cluster stability.

To restrict the rotational analysis to configurations with coherent
structure, in the plots of $\langle\omega\rangle$ versus $F^D$ in
Fig.~\ref{fig4}(c) and $G_6$ versus $F^D$ in Fig.~\ref{fig4}(d), we
only display points where $\mathrm{RMSD}\leq0.5$.
We use this threshold as an operational criterion for cluster
cohesion.
Within this restricted set of configurations, the angular response is
strongly dependent on the obstacle spacing.
The most pronounced negative excursion of $\langle\omega\rangle$
occurs for $a=1.00$ at low drive, whereas other spacings produce
weaker negative excursions or $\langle\omega\rangle\approx0$.
A smaller positive response is visible for $a=1.50$ over an
intermediate drive interval.
Thus, both the magnitude and sign of the rotational response can change
with the obstacle-site spacing and applied drive.

In Fig.~\ref{fig4}(d), the $G_6$ curve for $a=1.50$ takes a small
positive value of $G_6\approx0.05$, corresponding to approximately
one $60^{\circ}$ counterclockwise turn for every 20 obstacle sites.
For $a=1.00$, the pronounced negative response reaches values close to
$G_6=-0.5$, while several other curves remain near zero.
Taken together, the four observables in Fig.~\ref{fig4} show that
changing $a$ modifies the transverse transport, structural distortion,
and rotation in distinct ways, so a joint analysis is necessary to
identify the drive intervals in which an appreciable rotational
response coexists with a small RMSD.

\begin{figure}[t]
  \centering
  \includegraphics[width=1.0\linewidth]{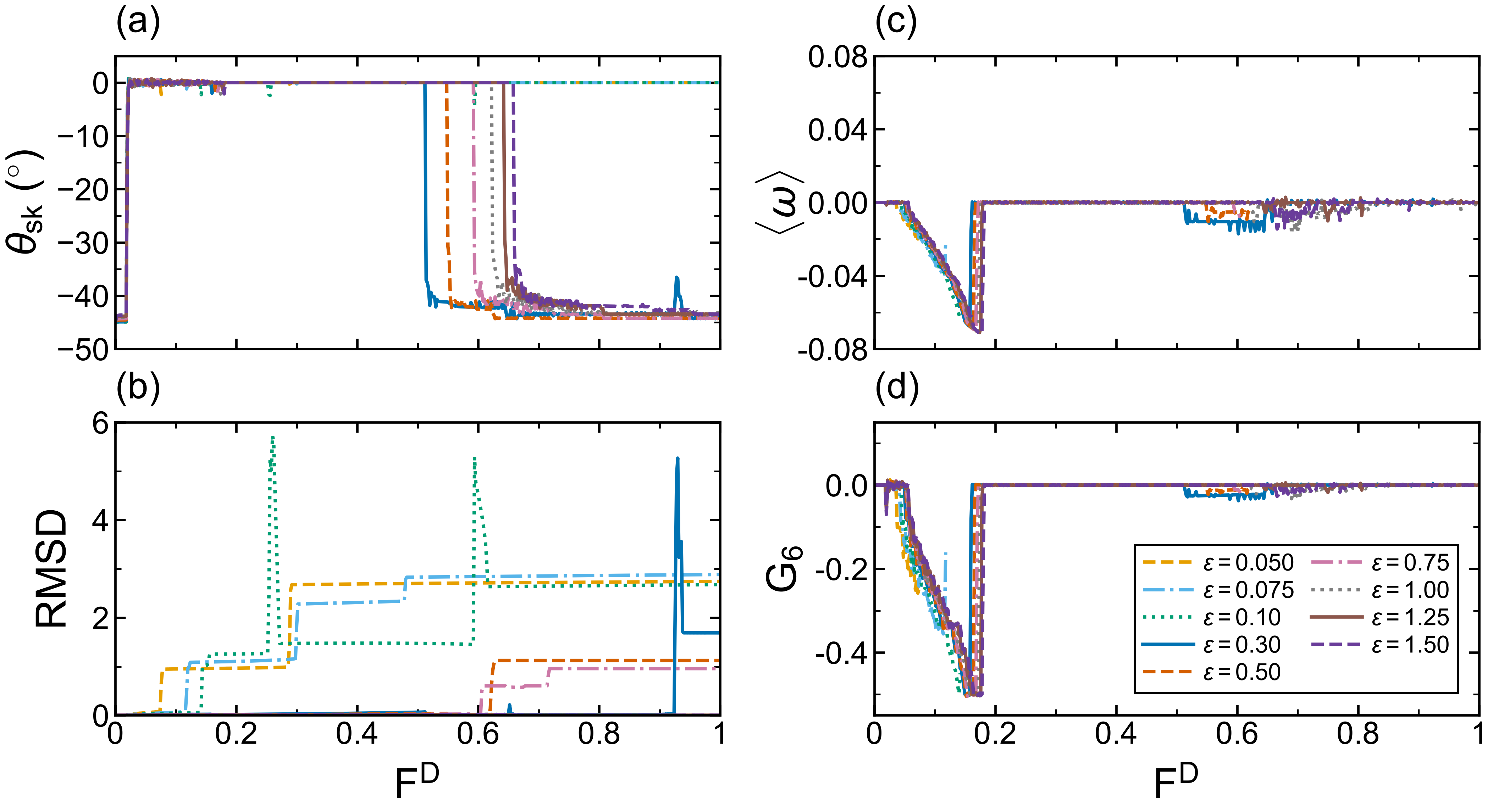}
  \caption{
  Dependence of the cluster dynamics on the interaction strength
$\epsilon$ for fixed obstacle spacing $a=1.0$.
(a) Skyrmion Hall angle $\theta_{\rm sk}$ versus $F^D$.
(b) Root-mean-square deviation (RMSD) from the relaxed cluster
configuration versus $F^D$.
(c) Average angular velocity $\langle\omega\rangle$ versus $F^D$.
(d) Gear ratio $G_6$ versus $F^D$.
The different curves correspond to
$\epsilon=0.05$ (gold dashed),
$0.075$ (light-blue dash-dotted),
$0.10$ (green dotted),
$0.30$ (blue solid),
$0.50$ (orange dashed),
$0.75$ (magenta dash-dotted),
$1.00$ (gray dotted),
$1.25$ (brown solid), and
$1.50$ (purple dashed).
In panels (c) and (d), only data with
$\mathrm{RMSD}\leq0.5$ are shown, so that the rotational response is
restricted to regimes in which the cluster remains approximately intact.}
  \label{fig5}
\end{figure}

Next, we examine the role of the interaction strength $\epsilon$
between the skyrmions.
We fix the obstacle spacing to $a=1.0$, at which our representative
system exhibits a pronounced low-drive rotational response, and vary
$\epsilon$ over a broad range.
In Fig.~\ref{fig5}(a--d) we plot the Hall angle
$\theta_{\mathrm{sk}}$, RMSD, mean angular velocity
$\langle\omega\rangle$, and gear ratio $G_6$ versus $F^D$ for
different values of $\epsilon$.

The structural response, shown as RMSD in Fig.~\ref{fig5}(b), exhibits
the largest change as $\epsilon$ is varied.
For the weakest interactions, the RMSD first increases from zero at
relatively low drives and then develops finite plateaus, indicating
that the cluster is readily deformed by its interaction with the
obstacle array.
As $\epsilon$ is increased, the cluster retains its reference
structure over a wider range of drives.
This behavior is consistent with the increased cohesion of the cluster
expected for stronger attractive interactions.

The Hall response in Fig.~\ref{fig5}(a) also changes with $\epsilon$.
For the cases that remain structurally coherent, the transition from
guided motion with $\theta_{\rm sk}\simeq0$ to motion with a finite
negative Hall angle shifts to higher values of $F^D$ as $\epsilon$
increases, showing that the ability of the cluster to cross the
obstacle barrier depends not only on the obstacle-site spacing but also
on the internal cohesiveness of the cluster.

The plots of $\langle\omega\rangle$ and $G_6$ versus $F^D$ in
Figs.~\ref{fig5}(c) and \ref{fig5}(d) show that the low-drive
rotational state persists over a broad range of interaction strengths.
For $F^D<0.2$ in Fig.~\ref{fig5}(c), a negative angular velocity
develops as the drive is increased, with the extent of the rotational
interval depending on $\epsilon$.
Correspondingly, in Fig.~\ref{fig5}(d), $G_6$ reaches negative values
of substantial magnitude and approaches $G_6\simeq-0.5$ for several
values of $\epsilon$, indicating a clockwise rotation close to one
$60^\circ$ turn for every two obstacle periods traversed by the
cluster.
The persistence of this response for different interaction strengths
indicates that the gear-like motion is robust over a wide range of
$\epsilon$.

The strong variation of the RMSD in Fig.~\ref{fig5}(b) demonstrates
that sufficient cluster cohesion is required for this rotational state
to remain well defined over an extended drive interval.
As in Fig.~\ref{fig4}, points with $\mathrm{RMSD}>0.5$ are omitted
from the rotational quantities in Fig.~\ref{fig5}(c,d) in order to
avoid interpreting the motion of strongly distorted configurations as
a coherent cluster rotation.

Overall, the rotational dynamics is considerably more sensitive to the
obstacle-site spacing $a$ than to the interaction strength $\epsilon$.
While $\epsilon$ mainly controls the cohesion and structural stability
of the cluster, allowing it to remain intact at higher drives, varying
$a$ produces stronger changes in the magnitude, sign, and extent of
the rotational response.
Thus, $a$ is a more effective control parameter for tuning distinct
gear-like dynamical states.

\subsection{Dynamical phase diagrams of obstacle parameters}
\label{subsec:phase-diagrams}

\begin{figure}[t]
  \centering
  \includegraphics[width=1.0\linewidth]{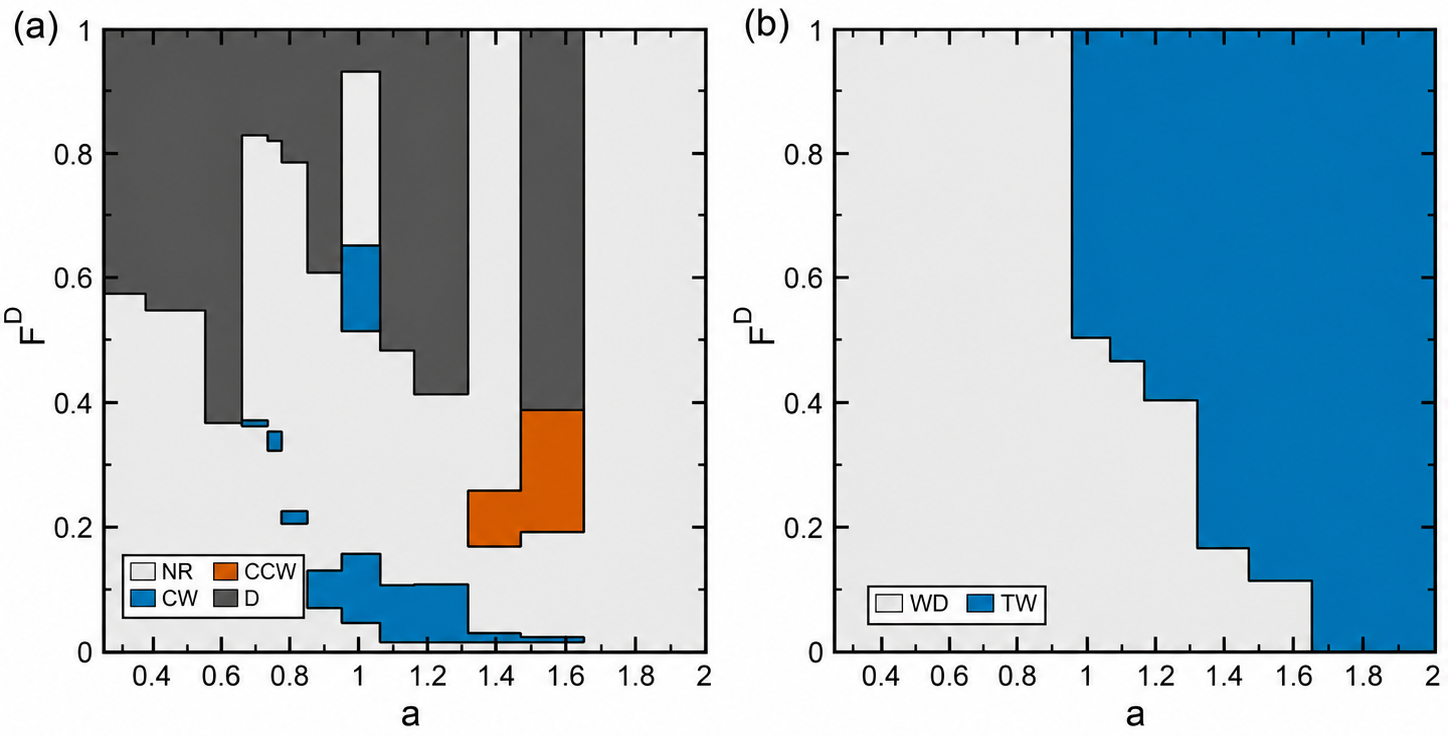}
  \caption{
  Dynamical phase diagrams in the drive--obstacle-spacing plane for
  fixed interaction strength $\epsilon=0.3$ and
  $\alpha_m/\alpha_d=1$.
  (a) Rotational and structural regimes: nonrotating (NR), clockwise
  rotating (CW), counterclockwise rotating (CCW), and deformed (D).
  The D regime corresponds to configurations with
  $\mathrm{RMSD}>0.5$.
  (b) Transport regimes relative to the obstacle barrier:
  wall dynamics (WD), in which the cluster remains confined to one
  side of the obstacle array, and traverse-wall dynamics (TW), in which
  the cluster crosses the barrier.
  In each panel, boundaries along $a$ are placed midway between
  consecutive simulated obstacle spacings.}
  \label{fig6}
\end{figure}

To summarize the different dynamical regimes identified above, we
construct dynamical phase diagrams as a function of $F^D$ versus $a$
in Fig.~\ref{fig6} for fixed interaction strength $\epsilon=0.3$.
For Fig.~\ref{fig6}(a), we classify the dynamics as nonrotating (NR),
clockwise rotating (CW), counterclockwise rotating (CCW), or deformed
(D).
The deformed regime corresponds to configurations that no longer
satisfy the cohesion criterion $\mathrm{RMSD}\leq0.5$ used to preserve
a well-defined coherent cluster for the rotational analysis.
Since the simulations are performed at discrete values of $a$, the
boundaries along the horizontal axis are placed midway between
consecutive simulated obstacle spacings.

The phase diagram in Fig.~\ref{fig6}(a) shows that coherent rotation
occupies restricted regions of parameter space rather than occurring
generically whenever the cluster interacts with the obstacle array.
Clockwise rotation is favored at intermediate obstacle spacings and is
most pronounced near $a=1.0$, where both the low-drive rotational
regime and a second rotating interval at higher drive are present.
At larger spacings, near $a=1.4$--$1.5$, an intermediate-drive CCW
regime appears, demonstrating that the obstacle geometry can also
reverse the sense of collective rotation.
In contrast, sufficiently small or large spacings predominantly
produce nonrotating motion, while deformation occupies an increasingly
large portion of the high-drive regime for several intermediate values
of $a$.

In Fig.~\ref{fig6}(b), we classify the same $F^D$--$a$ region
according to whether the cluster is guided by the obstacle wall and
remains on one side of it (wall dynamics, WD), or crosses the wall
(traverse-wall dynamics, TW).
In the wall-dynamics regime, the cluster remains locked at
$\theta_{\rm sk}=0^{\circ}$ and moves along the wall, whereas in the
traverse-wall regime the cluster passes through the obstacle barrier.
For closely spaced obstacle sites, the cluster remains in the WD regime
over the entire range of $F^D$ explored.
As $a$ increases, the drive required to cross the barrier decreases,
consistent with the larger openings between neighboring obstacle sites.
For the largest spacings, the barrier becomes sufficiently permeable
that traverse-wall motion occurs over essentially the entire drive
interval.

A comparison of Figs.~\ref{fig6}(a) and \ref{fig6}(b) reveals two
types of rotational behavior.
The low-drive CW regime occurs predominantly while the cluster is
guided along the obstacle wall, whereas the CCW state found near
$a=1.4$--$1.5$ appears only after the cluster has entered the
traverse-wall regime.
Thus, collective rotation can occur both during guided motion and
after transmission through the obstacle array, with the sense of
rotation depending on the dynamical regime.

\subsection{Influence of the Magnus term}
\label{subsec:magnus}

We next examine the influence of the Magnus term by fixing
$a=1.0$ and $\epsilon=0.3$ and varying the ratio
$\alpha_m/\alpha_d$ while maintaining
$\alpha_d^2+\alpha_m^2=1$.
Changing $\alpha_m/\alpha_d$ also changes the intrinsic skyrmion Hall
angle \cite{Lin13}, $\theta_{\rm sk}^{\rm int}
=
-\tan^{-1}\left(\frac{\alpha_m}{\alpha_d}\right).$
If the drive direction were kept fixed, the cluster would therefore
approach the obstacle wall at different angles for different
$\alpha_m/\alpha_d$.
To isolate the effect of the Magnus term, we adjust the drive angle
$\phi_D$ according to
$\phi_D
=
-45^\circ
+
\tan^{-1}\left(\frac{\alpha_m}{\alpha_d}\right),$
so that the free-flow velocity remains oriented at $-45^\circ$ for
all values of $\alpha_m/\alpha_d$.
Thus, the incidence direction and free-flow speed are kept fixed while
the relative Magnus contribution is varied.

\begin{figure}[t]
  \centering
  \includegraphics[width=0.7\linewidth]{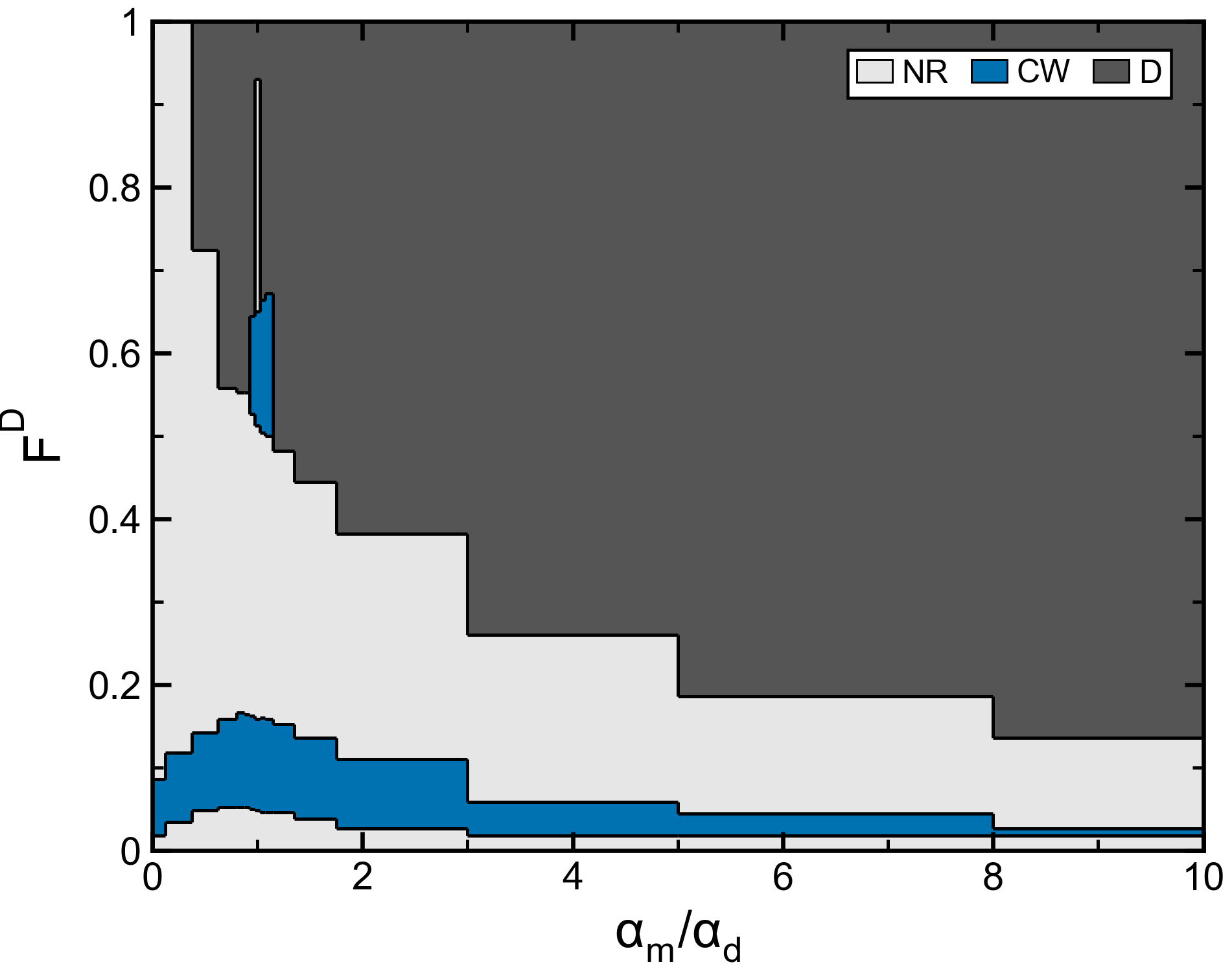}
  \caption{
  Dynamical phase diagram as a function of applied drive $F^D$ and
  Magnus-to-damping ratio $\alpha_m/\alpha_d$ for $a=1.0$ and
  $\epsilon=0.3$.
  The phases are nonrotating (NR), clockwise rotating (CW), and
  deformed (D).
  For each value of $\alpha_m/\alpha_d$, the drive direction is chosen
  so that the free-flow velocity remains oriented at $-45^\circ$.
  Boundaries along $\alpha_m/\alpha_d$ are placed midway between
  consecutive simulated ratios.}
  \label{fig7}
\end{figure}

Figure~\ref{fig7} shows the resulting dynamical phase diagram as a
function of $F^D$ versus $\alpha_m/\alpha_d$.
A low-drive CW regime persists over the full range of
$\alpha_m/\alpha_d$, including the purely dissipative limit
$\alpha_m/\alpha_d=0$, indicating that the Magnus term is not required
to convert translational motion into rotational motion.
Increasing $\alpha_m/\alpha_d$ from zero can broaden the CW rotational
regime and, for intermediate values of $\alpha_m/\alpha_d$, produces a
second CW regime at higher drives.
At the same time, the Magnus term reduces the structural stability of
the cluster, shifting the onset of the deformed regime toward lower
drives as $\alpha_m/\alpha_d$ increases.
This competition between enhanced rotational motion and reduced
cluster stability is clearly visible in Fig.~\ref{fig7}.

\section{Discussion}
\label{sec:discussion}

In this work, we considered a cluster of $N=7$ particles, which is the
smallest compact cluster consisting of a central particle surrounded
by a complete six-particle shell.
Other ordered cluster sizes should also be possible, such as the next
complete hexagonal shell at $N=19$, while for some choices of $N$ the
cluster may become partially disordered if the particles are unable to
arrange themselves in evenly spaced shells around a central particle.
Previous work on cluster ordering as a function of particle number has
identified different ordered and disordered cluster configurations
\cite{Bedanov94}.
For particle numbers where the cluster is disordered, enhanced sliding
rather than coherent gear-like rotation may occur.

We have focused on a line of repulsive defects, but rotational motion of a
cluster may arise more generally when there is a spatial gradient or
structured variation in the substrate pattern.
Pinning arrays with gradients have been studied and experimentally
realized in superconducting systems \cite{Ray13,Wang13,Lyu21}, and
related geometries should be realizable for skyrmion systems.
We have focused on skyrmions because driven skyrmion clusters have
already been observed experimentally \cite{Yu20,Wang22}; however, our
results suggest that related dynamics could also arise for other
cluster-forming systems, such as colloidal clusters moving over
one-dimensional periodic substrates.
There are also active-matter realizations with cluster geometries
similar to those considered here, in which the clusters themselves
rotate due to activity \cite{Tasci16,Aubret18}.
It would be interesting to determine how such active clusters interact
with a line defect and whether a transition occurs from predominantly
rotational motion to coupled rotational and translational motion.

Our results identify several dynamical states of a driven skyrmion
cluster interacting with a line of periodic repulsive obstacle sites.
When the drive is parallel to the obstacle line, the skyrmion Hall
effect pushes the cluster toward the substrate, producing a rotating
or gear-like phase in which particles at the edge of the cluster are
temporarily trapped by the obstacle sites.
We also observe a sliding phase in which the cluster moves along the
obstacle row without net rotation, a traverse-wall state in which it
crosses the obstacle line, and a strongly deformed state at high
drives.

These regimes are reflected in the velocity-force response, Hall
angle, root-mean-square deviation of the cluster, and dimensionless
gear ratio.
The rolling phase is most pronounced when the substrate spacing is
comparable to the characteristic interparticle spacing within the
cluster, while other obstacle spacings can produce weaker rotation or
counter-rotation.
The persistence of rotational states in the zero-Magnus limit
demonstrates that the basic translation-to-rotation conversion
mechanism does not rely on an intrinsic transverse force.
This broadens the relevance of the mechanism beyond skyrmions and
suggests that structured substrates could provide a general route for
controlling the rotational dynamics of driven particle clusters.

\section{Methods}
\label{sec:methods}

We simulate the dynamics of a seven-particle skyrmion cluster in a
$L_x\times L_y$ two-dimensional box with periodic boundary conditions
in both the $x$ and $y$ directions.
We use $L_x=L_y=36$ and consider a compact cluster of $N=7$ skyrmions
composed of one central skyrmion surrounded by six peripheral
skyrmions in a hexagonal arrangement, as illustrated in
Fig.~\ref{fig1}.
A one-dimensional periodic array of repulsive Gaussian obstacle sites
is placed along $y=2.0$, with a center-to-center spacing $a$ between
neighboring obstacle sites.

The skyrmion dynamics is governed by the particle-based equation of
motion \cite{Lin13}:
\begin{equation}
\alpha_d {\bf v}_i+
\alpha_m \hat{\bf z}\times{\bf v}_i
=
{\bf F}^{ss}_i+
{\bf F}^{p}_i+
{\bf F}^{D},
\label{eq:eom}
\end{equation}
where ${\bf v}_i=d{\bf r}_i/dt$ is the velocity of skyrmion $i$.
The first term on the left-hand side represents damping, while the
second term is the Magnus contribution, which generates a velocity
component perpendicular to the net force.
We normalize the coefficients such that
$\alpha_d^2+\alpha_m^2=1$ and, unless otherwise noted, use
$\alpha_m/\alpha_d=1$, giving
$\alpha_d=\alpha_m=1/\sqrt{2}$.

To model the attractive interaction between skyrmions expected in the conical phase of chiral magnets \cite{Leonov2016,Vizarim26,Kim2020}, we use an effective Lennard--Jones potential. This provides a qualitative first approximation that captures short-range repulsion and longer-range attraction, while neglecting deformations of the individual skyrmion textures and nonpairwise interactions. The potential is given by, 
\begin{equation}
U_{ss}(r_{ij})
=
4\epsilon
\left[
\left(\frac{\sigma}{r_{ij}}\right)^{12}
-
\left(\frac{\sigma}{r_{ij}}\right)^6
\right],
\label{eq:LJ}
\end{equation}
where $r_{ij}=|{\bf r}_i-{\bf r}_j|$ is the distance between
skyrmions $i$ and $j$, $\epsilon$ controls the interaction strength,
and $\sigma$ determines the characteristic length scale.
The skyrmion--skyrmion force is
${\bf F}^{ss}_i=-\sum_{j\neq i}\nabla_i U_{ss}(r_{ij})$.
We fix $\sigma=1.5$ and truncate the interaction at
$r_{ij}=2\sigma$.
Unless otherwise indicated, we use $\epsilon=0.3$.

Each obstacle site is represented by a repulsive Gaussian potential,
\[
U_p(r_{ip})
=
U_p^0
\exp\left[
-\left(\frac{r_{ip}}{r_p}\right)^2
\right],
\]
where $r_{ip}=|{\bf r}_i-{\bf r}_p|$ is the distance between skyrmion
$i$ and obstacle site $p$.
The corresponding force is
${\bf F}_{i}^{p}=-\nabla U_p$.
We use $U_p^0=1.0$ and $r_p=0.3$ throughout this work and vary the
obstacle lattice spacing $a$.

Unless otherwise noted, the external dc drive is applied along the
positive $x$ direction,
${\bf F}^{D}=F^{D}\hat{\bf x}$.
The equations of motion are integrated using a second-order
Runge--Kutta method with a time step $\Delta t=0.001$ at zero
temperature.
Before applying the drive, the cluster is relaxed for
$5\times10^5$ simulation steps.
We then increase $F^{D}$ in increments of
$\delta F^{D}=0.002$ and spend $2\times10^5$ simulation steps at each
drive value.

We measure the average longitudinal and transverse velocities,
$\langle V_x\rangle$ and $\langle V_y\rangle$, and the corresponding
Hall angle,
\[
\theta_{\rm sk}
=
\tan^{-1}
\left(
\frac{\langle V_y\rangle}{\langle V_x\rangle}
\right).
\]

To characterize the collective rotation of the cluster, we determine
the rotation that provides the best least-squares alignment between
the relaxed reference configuration and the instantaneous particle
configuration \cite{umeyama1991,lawrence2019}.
We define the geometrical cluster centers as
${\bf r}_c=N^{-1}\sum_i{\bf r}_i$ and
${\bf r}_c^0=N^{-1}\sum_i{\bf r}_i^0$, and introduce the relative
coordinates
${\bf p}_i={\bf r}_i-{\bf r}_c$ and
${\bf q}_i={\bf r}_i^0-{\bf r}_c^0$.
In two dimensions, minimization of the least-squares alignment error
gives the collective orientation
\begin{equation}
\theta =
{\rm atan2}
\left[
\sum_i ({\bf q}_i\times{\bf p}_i)_z,
\sum_i {\bf q}_i\cdot{\bf p}_i
\right],
\label{eq:theta}
\end{equation}
where $\theta\in(-\pi,\pi]$.

We remove the $2\pi$ discontinuities between successive measurements
to construct the continuous unwrapped angle $\theta_{\rm unw}$, from
which the angular velocity is obtained as
$\omega=d\theta_{\rm unw}/dt$.
When the cluster crosses a periodic boundary, the particle coordinates
are reconstructed using the minimum-image convention before evaluating
the cluster center and rotational quantities.

To quantify deviations of the cluster from its relaxed structure, we
calculate the root-mean-square deviation (RMSD) after removal of the
global translation and optimal rigid-body rotation
\cite{coutsias2004,hanson2020},
\begin{equation}
{\rm RMSD}
=
\left[
\frac{1}{N}
\sum_i
\left|
{\bf p}_i-R(\theta){\bf q}_i
\right|^2
\right]^{1/2},
\label{eq:rmsd}
\end{equation}
where $R(\theta)$ is the two-dimensional rotation matrix associated
with the best-fit orientation defined in Eq.~(\ref{eq:theta}).
Since global translation and rotation have been removed, the RMSD
quantifies internal deviations from the relaxed cluster configuration,
with increasing values indicating stronger structural deformation.

Finally, since the six peripheral particles have an approximately
sixfold symmetry, we define the dimensionless gear ratio
\begin{equation}
G_6=
\frac{6a\langle\omega\rangle}
{2\pi\langle V_x\rangle},
\label{eq:G6}
\end{equation}
which measures the average cluster rotation, in units of $60^\circ$,
per obstacle lattice period traveled along the $x$ direction.
The sign of $G_6$ distinguishes the two senses of rotation.


\section*{Data availability}

The data generated and analyzed during the current study are available
from the corresponding author upon reasonable request.

\bibliography{refs}

\section*{Acknowledgements}

This work was supported by the U.S. Department of Energy through Los
Alamos National Laboratory.
Los Alamos National Laboratory is operated by Triad National Security,
LLC, for the National Nuclear Security Administration of the U.S.
Department of Energy (Contract No. 892333218NCA000001).

N.P.V. acknowledges computational resources funded by the
Funda\c{c}\~ao de Amparo \`a Pesquisa do Estado de S\~ao Paulo
(FAPESP), Grant No. 2024/02941-1.

\section*{Author contributions}

N.P.V. performed the simulations and data analysis and prepared the
initial manuscript.
C.J.O.R. and C.R. contributed to the conception of the study,
interpretation of the results, and revision of the manuscript.
All authors reviewed and approved the final manuscript.

\section*{Competing interests}

The authors declare no competing interests.

\end{document}